# Mesoscopic mechanism of the domain wall interaction with elastic defects in ferroelectrics

Anna N. Morozovska[1*], Eugene A. Eliseev[1,2], G.S. Svechnikov[1], and Sergei V. Kalinin[3†]

[1] Institute of Semiconductor Physics, National Academy of Science of Ukraine,
41, pr. Nauki, 03028 Kiev, Ukraine

[2] Institute for Problems of Materials Science, National Academy of Science of Ukraine,
3, Krjijanovskogo, 03142 Kiev, Ukraine

[3] Center for Nanophase Materials Science, Oak Ridge National Laboratory,
Oak Ridge, TN, 37831

**Abstract**

The role of elastic defects on the kinetics of 180-degree uncharged ferroelectric domain wall motion is explored using continuum time-dependent LGD equation with elastic dipole coupling. In one dimensional case, ripples, steps and oscillations of the domain wall velocity appear due to the wall-defect interactions. While the defects do not affect the limiting-wall velocity vs. field dependence, they result in the minimal threshold field required to activate the wall motions. The analytical expressions for the threshold field are derived and the latter is shown to be much smaller than the thermodynamic coercive field. The threshold field is linearly proportional to the concentration of defects and non-monotonically depends on the average distance between them. The obtained results provide the insight into the mesoscopic mechanism of the domain wall pinning by elastic defects in ferroelectrics.

[*] morozo@i.com.ua
[†] sergei2@ornl.gov

# 1. Introduction

Domain walls dynamics in ferroelectric materials is the key factor underpinning kinetics of polarization switching of ferroelectric materials and devices [1], and hence directly relevant to operation of ferroelectric memory devices. Furthermore, domain wall dynamics underpins many unique properties of polycrystalline and disordered ferroelectrics, including enhanced piezoelectric properties and ferroelectric nonlinearity [2,3]. Domain walls in ferroelectric materials have attracted much attention in the context of physical phenomena enabled by charge accumulation and associated band bending [4, 5, 6], ferroelastic transitions [7], band gap narrowing [8], impurity segregation [9, 10] and order parameter couplings [11, 12, 13]. Functional properties of domain walls have become accessible for direct observations through scanning probe microscopy methods [7, 14, 15, 16], whereas associated structures including order parameter components are now amenable to atomic resolution electron microscopy techniques [17, 18].

Many aspects of polarization dynamics in ferroelectric are intrinsically linked to domain wall pinning on structural defects. These include point defects including vacancy and impurity atoms, defect clusters, linear defects such as dislocations, and planar defects such as antiphase boundaries, Ruddlesden-Popper interfaces, ordered oxygen vacancies, and domain walls corresponding to other order parameters [19]. These interactions lead to well-known phenomena such as aging and remnant polarization degradation. Similarly, recent detailed studies of conductivity at the walls have demonstrated an intriguing set of hysteretic and memory effects and time dynamics [20], suggesting that this dynamics is strongly affected by domain wall pinning on defects.

Analysis of the domain wall kinetics requires detailed understanding of the domain wall pinning mechanisms in the presence of defect centres. Whereas the mechanisms of pinning by charged defects and their influence on polarization dynamics are well-studied [21, 22, 23, 24, 25], this is not the case for elastic (structural) defects, even though there are many defects of such type including uncharged vacancies, aliovalent atoms, and extended defects self-ordered in chains, planes, anti-distortive boundaries and elastic twins. Here, we consider the mesoscopic mechanisms of uncharged domain wall pinning by planar elastic defects and analyse defect-wall interactions. This analysis can be easily extended to the 3D defect structure by well-elaborated statistical methods.

## 2. Statement of the problem

For a domain wall propagating in *x*-direction in the second order ferroelectric, the spontaneous polarization component $P_z$ distribution obeys time-dependent LGD equation:

$$\Gamma\frac{\partial P_z}{\partial t} + \alpha(T,\mathbf{r})P_z + \beta P_z^3 - g\left(\frac{\partial^2 P_z}{\partial x^2} + \frac{\partial^2 P_z}{\partial y^2} + \frac{\partial^2 P_z}{\partial z^2}\right) = E_0 + E_d(\mathbf{r}) \tag{1a}$$

External field $E_0$ is applied along z-axes at times $t>0$. Depolarization field $E_d(\mathbf{r}) = -\frac{\partial \varphi}{\partial z}$ can be found from the Debye equation $\Delta\varphi - \frac{\varphi}{R_d^2} = \frac{\partial P_z}{\varepsilon_0\varepsilon_b \partial z}$. In the bulk of ferroelectric depolarization field is expressed as [26]:

$$E_d(\mathbf{r}) = -\int_{\mathbf{k}} d\mathbf{k}\, \frac{k_z^2 \widetilde{P}_z(\mathbf{k})\exp(-i\mathbf{kr})}{\varepsilon_0\varepsilon_b\left(k^2 + R_D^{-2}\right)} \tag{1b}$$

where $R_D$ is a Debye screening radius; $\varepsilon_b \sim 10$ is the relative dielectric constant of background, $\varepsilon_0 = 8.85\times10^{-12}$ F/m is the dielectric permittivity of vacuum. $E_d(\mathbf{r})$ vanishes for uncharged 180-degree *x*-domain wall in the space regions very far from defects, where $P_z \approx P_z(x)$, but in the vicinity of defects it may be noticeable, making integro-differential Eq.(1a) extremely complex for analytical treatment due the nonlocality of depolarization field (1b).

Note, that tilted domain walls have additional positive depolarization field energy due to the bound charge and thus have higher energy then uncharged ones. Considering here the case of uncharged 180-degree walls we also minimize the influence of depolarization field and screening on the domain wall kinetics and can explore the "net" effects originated from elastic defects. Moreover, when the wall is pinned at the defect, the most stable state is when it is flat and thus uncharged. Estimations of the depolarization field strength caused by the interaction of uncharged domain wall with well-localized elastic fields in the ferroelectric material with free carriers are made in **Appendix A1.**

The microscopic origins of elastic defects include impurity atoms and dilatation centers, which can be either point or form clusters, planar structures as trapped by anti-distortive boundaries. These defects couple into the free energy via the elastic dipole (Vegard) mechanism (see **Appendix A2** and Ref.[27, 28, 29]). Allowing for elastic dipole coupling, the coefficient $\alpha(T,\mathbf{r})$ depends on the elastic stress $\sigma_{ij}(\mathbf{r})$ created by dilatation centers (e.g. impurity atoms, oxygen vacancies) local concentration as:

$$\alpha(T,\mathbf{r}) = a(T) + 2Q_{33ij}\sigma_{ij}(\mathbf{r}), \qquad \sigma_{ij}(\mathbf{r}) = \beta_{ij}^d\left(N_d(\mathbf{r}) - \overline{N}_d\right). \tag{2}$$

Here $a(T) = \alpha_T (T - T_C)$, $Q_{33ij}$ is the electrostriction tensor component, $\beta_{ij}^d$ is the elastic dipole tensor (equivalent name is Vegard molar expansion tensor). $N_d(\mathbf{r})$ is the local concentration of *elastic defects*, some of which can be neutral or charged: $N_d = N_d^0 + N_d^+ + N_d^-$; $\overline{N}_d$ is their average concentration. Note that nearly always $N_d^\pm << N_d$ [30], because in equilibrium the probability of a static defect to be ionized is much smaller than unity; while ionized defects are typically well-screened by free carriers. The fact allows us not to consider charged defects in Eq.(2) and use the approximation $\sigma_{ij}(\mathbf{r}) \approx \beta_{ij}^d (N_d^0 - \overline{N}_d)$, where only elastic defects are considered. Since the term $2Q_{33ij}\sigma_{ij}(\mathbf{r})$ formally renormalizes Curie temperature $T_C$ in Eq.(2), the considered elastic defects can be classified as "random temperature"-type defects.

### 3. 1D model of uncharged wall interaction with planar elastic defects

Let us further consider a 1D-problem, when all quantities effectively depend only *on the distance x from the uncharged 180-degree domain wall*, located far from the sample boundaries (see **Fig. 1a**). The model situation can rigorously describe the wall interaction with planar defects in layered structures and elastic walls.

For the considered 1D-case polarization obeys equation $\Gamma \frac{\partial P_z}{\partial t} + (a(T) + 2Q_{33ij}\sigma_{ij}(x))P_z + \beta P_z^3 - g\frac{\partial^2 P_z}{\partial x^2} = E_0$ with initial condition as a single-domain wall non-perturbed by defects, i.e. $P(t=0, x) \approx P_S \tanh(x/\sqrt{2}R_c)$, and boundary conditions $\partial P_z(t, x \to \pm\infty)/\partial x = 0$ far from the wall, where *correlation length* $R_C = \sqrt{-g/a}$ determines the wall width, spontaneous polarization $P_S = \sqrt{-a/\beta}$. Depolarization field (1b) is absent for the 1D-case, since $P_z \equiv P_z(x)$. Estimations of the depolarization field (1b) performed in **Appendix A1** give a hope that the solution of the 1D-problem can explain qualitatively the main features of domain wall interaction with 3D-elastic defect fields in ferroelectric-semiconductors if the defects z-size becomes 5-10 times higher than the Debye screening radius.

Elastic field $\sigma_{ij}(x)$ created by defects assembles and single defects in semiconductors and dielectrics were calculated analytically [31, 32] within the framework of nonlinear continuum theory of elasticity. It was shown that elastic field structure depends on defect concentration and their elastic interaction; in particular defect self-ordering is possible with $\overline{N}_d$ increase. Concentration of elastic defects can be expanded in Fourier series in harmonic approximation and re-expanded in hyperbolic (exponential) functions allowing for elastic fields non-linearity. For the sake of definiteness of numerical simulations and analytical results obtained below, we will use 1D-periodic distribution of elastic stress (a) harmonic (e.g. sinusoidal) and (b) well-localized at each defect layer

$$\sigma_{ii}^d(x) = \sigma_d^i \cos(2\pi(x-x_0)/d), \tag{3a}$$

$$\sigma_{ii}^d(x) = \sigma_d^i \left( \begin{array}{c} \exp\left(-\ln(2)\dfrac{\cos(2\pi(x-x_0)/d)-1}{\cos(\pi w/d)-1}\right) \\ -\exp\left(\dfrac{-\ln(2)}{\cos(\pi w/d)-1}\right) I_0\left(\dfrac{\ln(2)}{\cos(\pi w/d)-1}\right) \end{array} \right), \tag{3b}$$

and (c) the field created by a single defect layer [31] for comparison:

$$\sigma_{ii}^d(x) \approx \frac{\sigma_d^i}{B + \cosh((x-x_0)/w)}. \tag{3c}$$

Here subscript $i = 2$ or 3, $\sigma_d^i = \beta_{ii}^d N_d^0$ is the amplitude, $d$ is the "main" period of the elastic field, related with the average distance between the defects as $d \sim \sqrt[3]{6/\pi N_d^0}$, $w$ is the width of the field distribution, $I_0$ is modified Bessel function of zero order. Distribution (3b) is shown in **Fig.1b**. Constant $|B| < 1$ defines a single defect strength in Eq.(3c) (see comments to Eq.(20) in Ref. [31]). Without losing the generality below we can consider the case $B \ll 1$. Distributions (3) have zero average value: $\dfrac{1}{L}\int_{-L}^{L} \sigma_{ii}^d(x)dx = 0$ at $L \to \infty$.

Considered problem is determined quantitatively by several parameters: time scale $\tau$, correlation length $R_C$, spontaneous polarization $P_S$, thermodynamic coercive field $E_C$, dimensionless strength $\Delta\alpha$ of elastic field created by defects, period $\delta$ (determined by the average distance between defects) and width $h$ defined as

$$\tau = \frac{\Gamma}{|a|}, \quad R_C = \sqrt{-\frac{g}{a}}, \quad P_S = \sqrt{-\frac{a}{\beta}}, \quad E_C = \sqrt{\frac{-4a^3}{27\beta}}, \quad \Delta\alpha = \frac{2Q_{33ij}\beta_{ij}^d N_d^0}{a(T)}, \quad \delta = \frac{d}{R_C}, \quad h = \frac{w}{R_C} \tag{4}$$

Estimation of the most important parameter $|\Delta\alpha|$ gives 0.02 – 0.1 for typical ferroelectric parameters $Q_{33ij} = (0.1–0.2)$ m$^4$/C$^2$, $\beta_{ij}^d = (3–30)$ eV [27], $N_d^0 = (10^{24}–10^{25})$m$^{-3}$ and $a(T) = 4\,10^7$m/F at $T = 300K$.

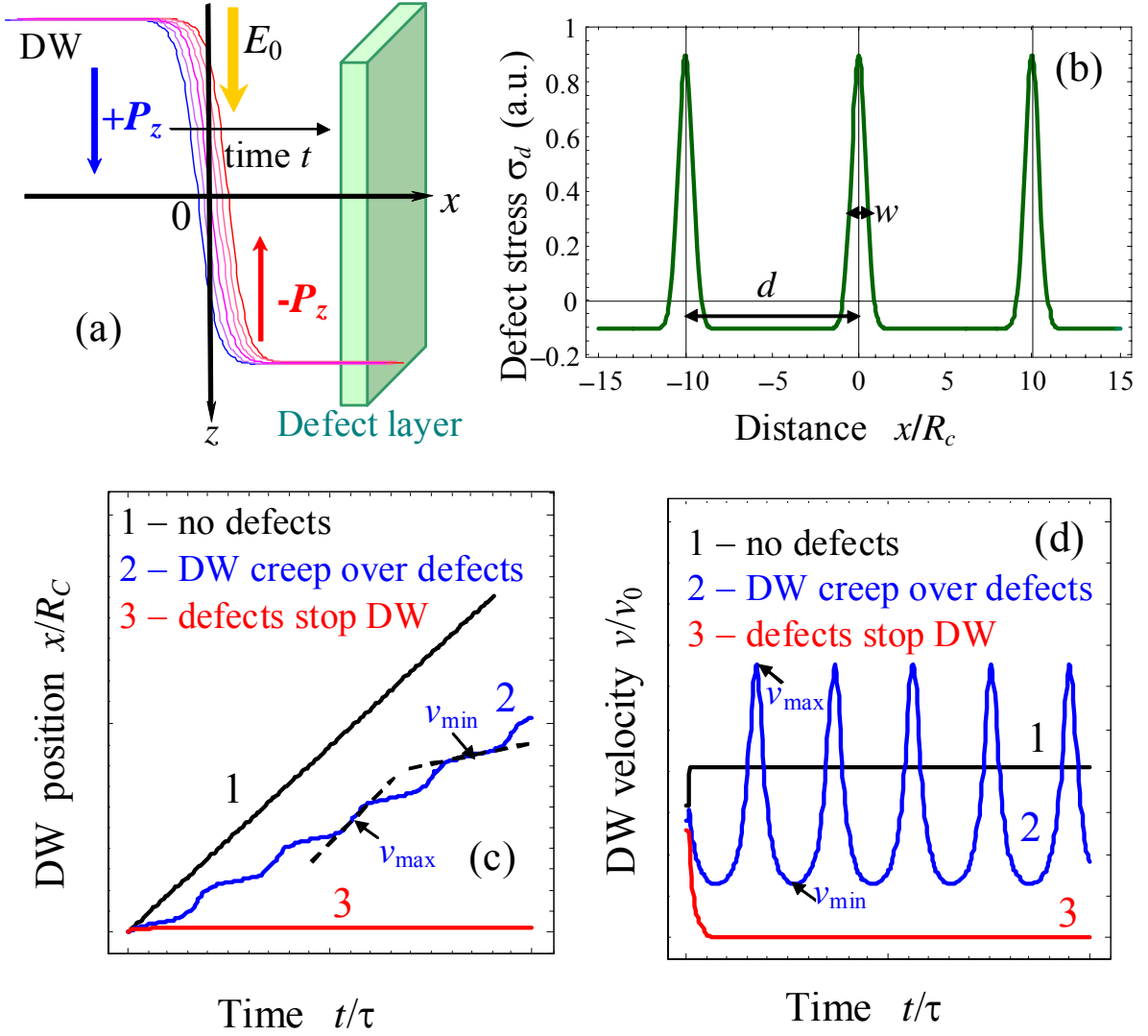

**Figure 1.** (a) Uncharged domain wall with polarization distribution $P_z(t,x)$ is moving in external field $E_0$. (b) Stress field $\sigma_{22}^d(x)$ created by elastic defects calculated from Eq.(3b) for $w/d=0.1$. (c, d) Domain wall position (c) and velocity (d) vs. time calculated for different concentration of elastic defect (curves 1-3) and the same external field $E_0$.

### 4. Results and discussion

Despite expressions (3) look rather specific, results of numerical simulations for domain wall kinetic appeared *qualitatively* insensitive to the concrete form of elastic field created by defects. Without elastic defects domain wall velocity $v$ is constant, proportional to the dragging field $E_0$ (see curves 1 in **Figs.1c,d**). Moderate amount/strength of elastic defects pin the domain wall at fixed field $E_0$: steps appear on the domain wall position and oscillations appear on the domain wall velocity due to the creeping on elastic defects (see curves 2 in **Figs.1c,d**). High amount/strength of elastic defects stop the domain wall at fixed field $E_0$ (see curves 3 in **Figs.1c,d**).

Details of the domain wall interaction with multiple defects are shown in **Figs.2-3.** Results are quantitatively sensitive to the width of defect layer $w$, layer period $d$ and amplitude $\sigma_d$ of the elastic field. One can compare domain wall position and velocity vs. time and dragging field calculated for harmonic distribution (3a), and more realistic well-localized anharmonic distribution (3b) from comparison of the plots in **Fig. 2-3**. Results for periodic layers are similar at the same $d$ and $\sigma_d$, but the difference in shape is related with the width $w$ value. Steps and oscillations appear due to the wall creeping over elastic defects. Domain wall stops for small dragging fields (see curves 1 and 2 in **Figs.2a,b** and curves 1 in **Figs.3a,b**), so the dragging threshold field $E_{th}$ appears due to the pinning on defects. The threshold field value is much smaller than the thermodynamic coercive field and depends on the elastic field distribution features (compare **Figs.2c** and **3c**). Minimal velocity of domain wall is equal to the difference $\sqrt{2/3}(E_0 - E_{th})/E_C$. At $E_0 >> E_{th}$ the slope of the minimal ($v_{\min}$), average ($v_{av}$) and maximal ($v_{\max}$) wall velocities dependences on the dragging electric field are the same as the slope of the wall velocity in the ferroelectric without defects (dashed and solid curves have the same slope). Namely for all the cases $v \sim \sqrt{2/3}\, E_0/E_C$ in accordance with Ishibashi [33].

Using direct variational method and harmonic approximation (3a) for elastic field, we derived an approximate analytical expression for the threshold field of the domain wall motion as:

$$\frac{E_{th}}{E_C} \approx \frac{3\sqrt{3}d^2\Delta\alpha}{2(d^2 + 2\pi^2 R_C^2)} \frac{R_C d}{2\sqrt{2}\pi R_C^2 - (5/4)R_C d + (3/2\sqrt{2}\pi)d^2} = \begin{cases} \dfrac{3\sqrt{3}\Delta\alpha}{\sqrt{2}}\left(\dfrac{d}{2\pi R_C}\right)^3, & d << 2\pi R_C \\ \dfrac{\sqrt{6}\Delta\alpha}{2}\left(\dfrac{2\pi R_C}{d}\right), & d >> 2\pi R_C \end{cases} \quad (5)$$

Dependence of the threshold field on the period $d$ is shown in **Fig.4a.** Detailed derivation of Eq.(5) is given in **Appendix A3**. Numerical calculations proved that the threshold field is proportional to $\Delta\alpha E_C$ and expression (5) is valid quantitatively for harmonic distribution (3a) and semi-quantitatively for an-harmonic distribution (3b) (compare triangles with dotted curve and other symbols in **Fig.3**). It follows from Eq.(5) that a threshold field is negligibly small for the case $d << R_C$ and vanishes as $1/d$ for the case $d >> R_C$. Threshold field $E_{th}/E_C$ vs. dimensionless defect concentration $R_C^3 N_d^0$ is shown in **Fig.4b**. Limiting expressions are valid:

$E_{th}/E_C \to \sqrt{3/2}(3/8\pi^3)\delta\tilde{\alpha}$ at $R_C^3 N_d^0 \gg 1$ and $E_{th}/E_C \to \sqrt{6}\pi(R_C^3 N_d^0)^{4/3}\delta\tilde{\alpha}$ at $R_C^3 N_d^0 \ll 1$, where $\delta\tilde{\alpha} = 2Q_{33ij}\beta_{ij}^d/(a(T)R_C^3)$.

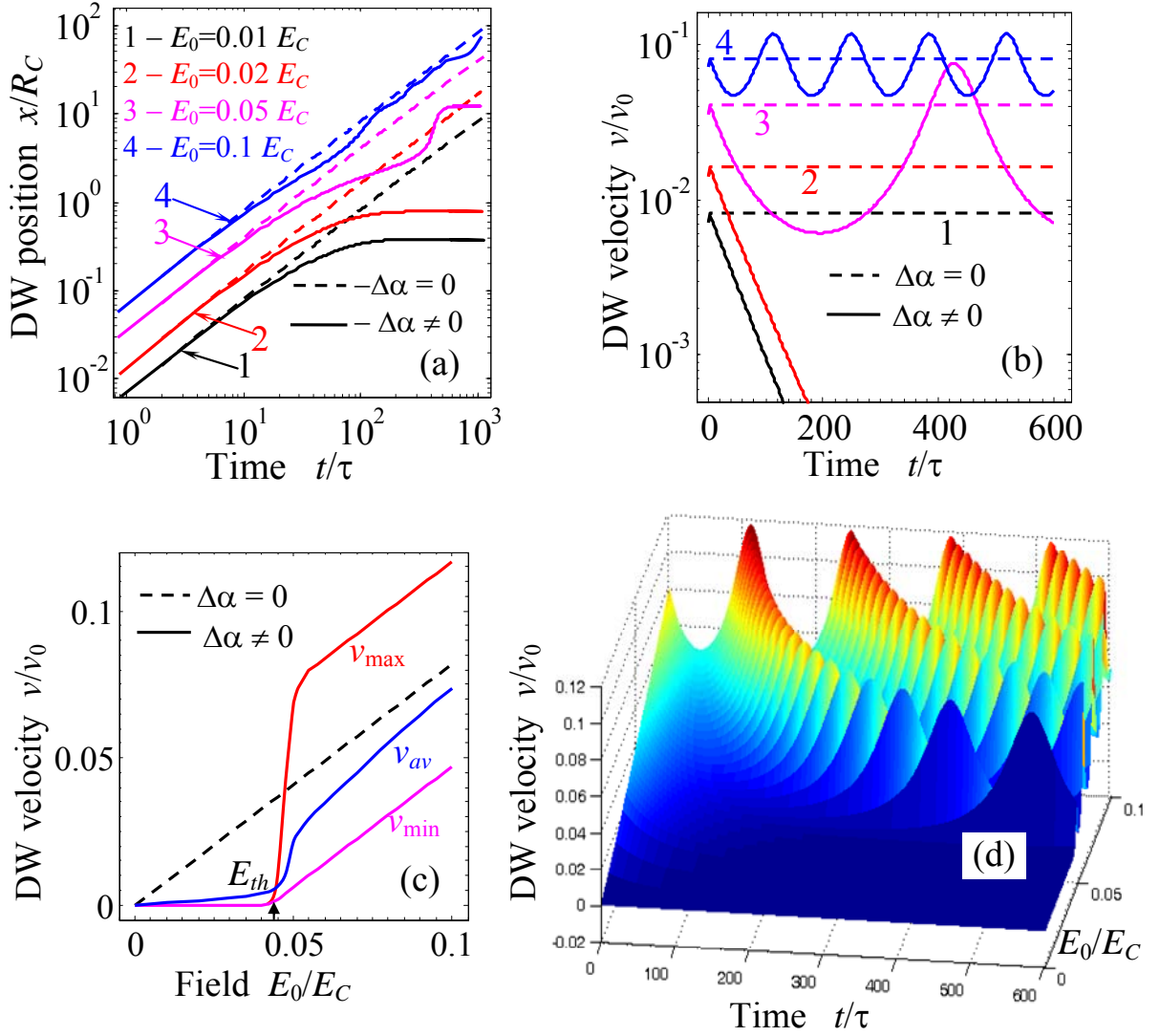

**Figure 2.** Domain wall position (a) and velocity (b) vs. time calculated for different dragging fields $E_0 = 0.01, 0.02, 0.05, 0.1 E_C$ (curves 1-4). Dashed curves are calculated for $\Delta\alpha = 0$ (defects are absent). Solid curves are calculated for ***harmonic modulation*** $\alpha(x) = a(1+\Delta\alpha\cos(2\pi x/d))$ followed from Eq.(3a) for defect elastic field for parameters $d = 10R_C$, $\Delta\alpha = 0.05$. Velocity scale $v_0 = R_C/\tau$. Ripples appear due to defects. (c) Corresponding domain wall minimum ($v_{min}$), average ($v_{av}$) and maximal ($v_{max}$) velocities $v/v_0$ vs. the dragging field $E_0/E_C$. (d) 3D plot of the domain wall velocity $v/v_0$ in coordinates dragging field $E_0/E_C$ and time $t/\tau$.

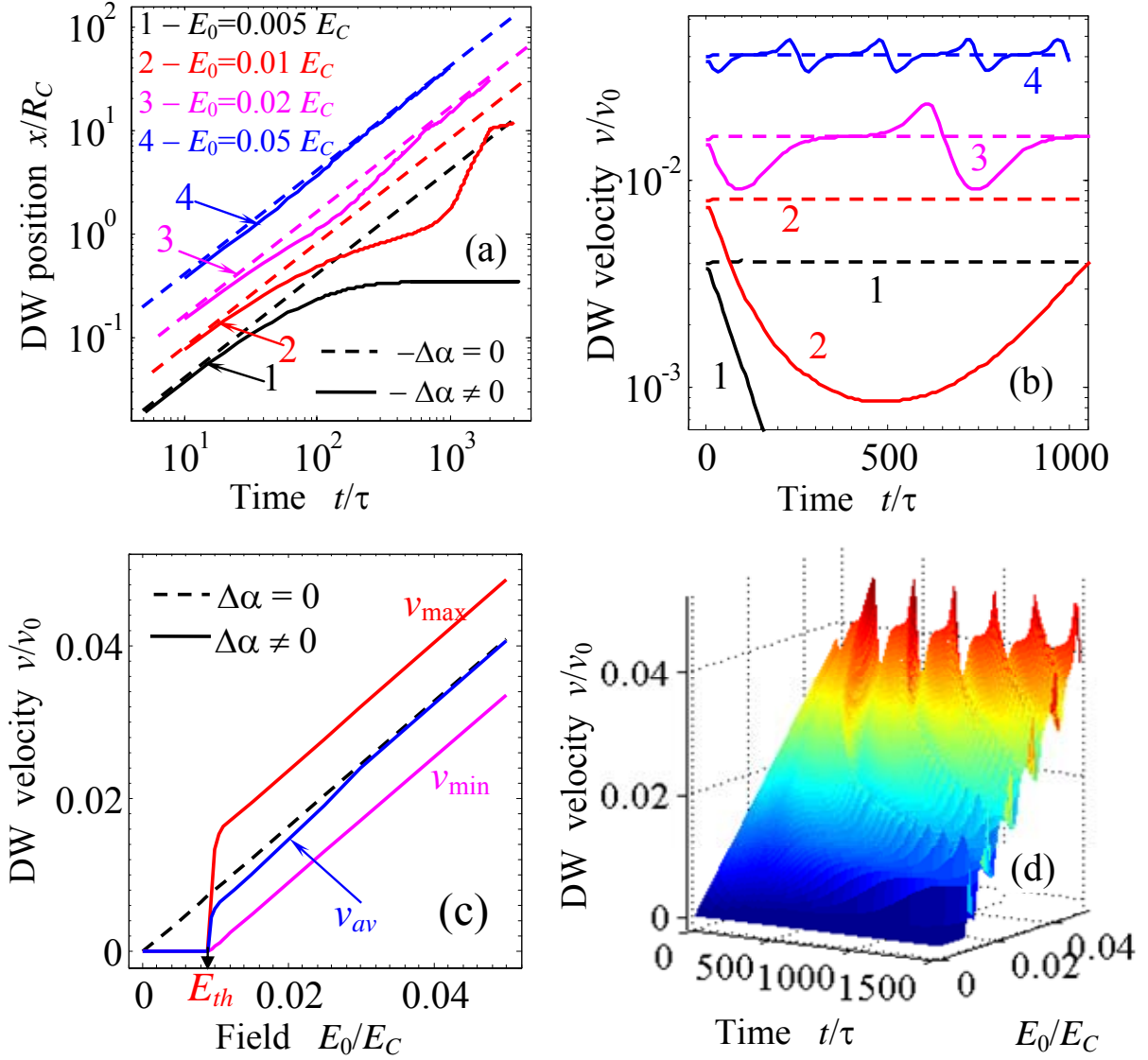

**Figure 3.** Domain wall position (a) and velocity (b) vs. time calculated for different dragging fields $E_0 = 0.005, 0.01, 0.02, 0.05 E_C$ (curves 1-4). Dashed curves are calculated for $\Delta\alpha = 0$ (defects are absent). Solid curves are calculated for **anharmonic modulation** $\alpha(x) = a(1 + \Delta\alpha f(2\pi x/d), w)$ from Eq.(3b), where $\Delta\alpha = 0.05$, period $d = 10 R_C$, width $w=0.05d$. Velocity scale $v_0 = R_C/\tau$. (c) Corresponding domain wall minimum ($v_{min}$), average ($v_{av}$) and maximal ($v_{max}$) velocities $v/v_0$ vs. the dragging field $E_0/E_C$. (d) 3D plot of the domain wall velocity $v/v_0$ in coordinates dragging field $E_0/E_C$ and time $t/\tau$.

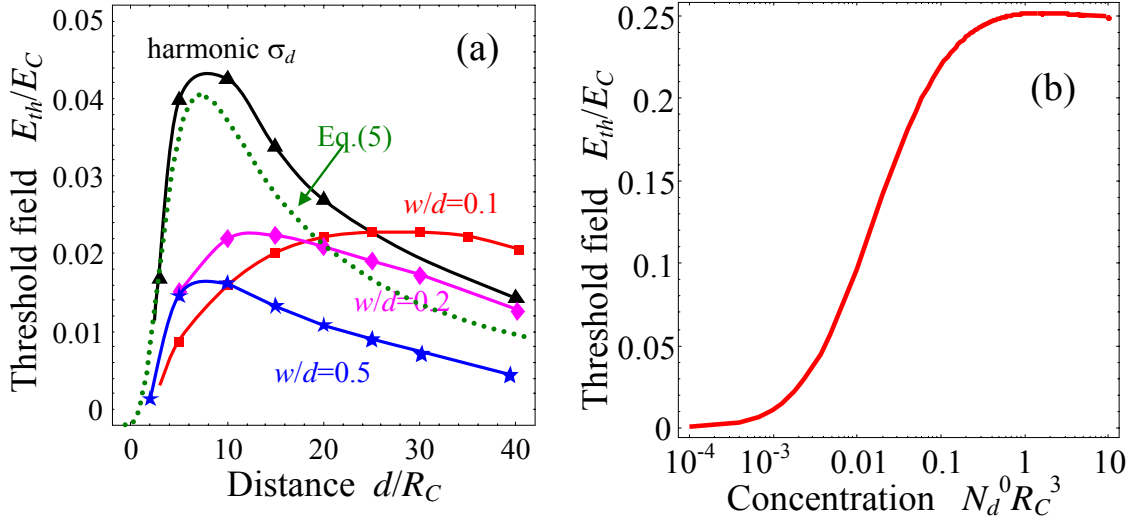

**Figure 4.** Dependence of the dimensionless threshold field $E_{th}/E_C$ on the relative average distance between elastic defects $\delta = d/R_C$. Triangles are calculated for harmonic elastic filed (3a), other symbols are calculated for anharmonic elastic field (3b) at different ratio $w/d$ : 0.1 (squires), 0.2 (rhombs), 0.5 (stars). Dotted curve is calculated from Eq.(5). Parameter $\Delta\alpha$ is fixed $\Delta\alpha = 0.05$. (b) Threshold field $E_{th}/E_C$ vs. dimensionless defect concentration $R_C^3 N_d^0$, for typical parameters $\delta\tilde{\alpha} = 16$ and $R_C = 0.5$ nm.

Details of the domain wall interaction with a single defect layer located at $x = 10R_C$ are shown in **Figs. 5.** Temporal evolution of the domain wall position and velocity was studied with increase of the dragging fields $E_0/E_C$ at the same defect layer thickness $w$ (see curves 1-4 in **Figs. 5a,b**) and with increase of $w$ at the same $E_0$ (see curves 1-4 in **Figs. 5c,d**). Defect layer stops if $E_0$ is less that the threshold value (see curves 1-2 in **Figs. 5a,b**). The threshold field of the domain wall motion is determined by the $w$ and $\Delta\alpha$. Unexpectedly we meet with the fact that the single defect layer of both very small and high thickness $w$ cannot stop the wall at fixed $E_0$ (compare curves 2 with curves 1, 3 and 4 in **Figs. 5c,d**). An approximate analytical dependence of the threshold field $E_{th}$ was derived as:

$$\frac{E_{th}}{E_C} \sim \frac{\sqrt{6\pi}\Delta\alpha w R_C}{(w^2 + R_C^2)} \sim \begin{cases} \Delta\alpha\left(\dfrac{w}{R_C}\right), & w \ll R_C \\ \Delta\alpha\left(\dfrac{R_C}{w}\right), & w \gg R_C \end{cases} \qquad (6)$$

The dependence (6) has maximum at $w/R_C = 1$. Derivation of Eq.(6) is similar to the derivation of Eq.(5).

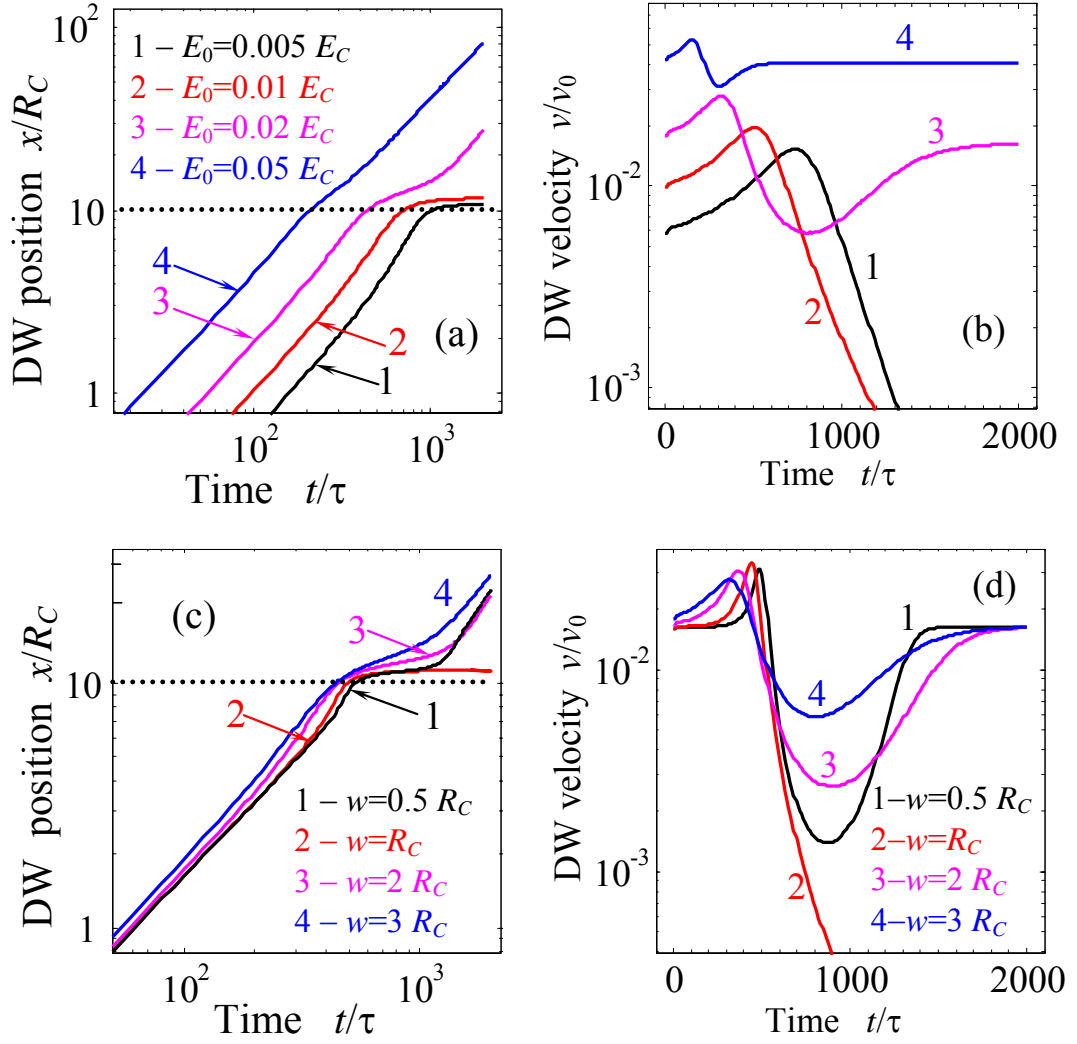

**Figure 5.** Domain wall position (a) and velocity (b) vs. time calculated for different dragging fields $E_0 = 0.005, 0.01, 0.02, 0.05 E_C$ (curves 1-4) and the same defect size $w = 3R_C$. Domain wall position (c) and velocity (d) vs. time calculated for different defect sizes $w/R_C = 0.5, 1, 2, 3$ (curves 1-4) and the same $E_0 = 0.02 E_C$. Curves are calculated for *a single planar defect (3c)* with parameter $\Delta\alpha = 0.05$ and center $x = 10 R_C$ (horizontal dotted line in a,c). Velocity scale $v_0 = R_C/\tau$.

**Summary**

Using continuum time-dependent LGD equation with elastic dipole (Vegard coupling) term we consider the influence of elastic defects on the kinetics of 180-degree uncharged ferroelectric domain wall. Ripples, steps and oscillations of the domain wall velocity appear due to the creeping on elastic defects. The slope of the minimal, average and maximal wall velocities dependences on the dragging electric field are the same as the slope of the wall velocity in the ferroelectric without defects, but the threshold field exists due to defects. We derived analytical expression for the threshold electric field required for the domain wall motion. The threshold field value is much smaller than the thermodynamic coercive field and is determined by the elastic field of defects. The threshold field is linearly proportional to the concentration of defects and non-monotonically depends on the average distance between them. We hope that obtained results could explore the mechanism of the domain wall pinning by elastic defects in ferroelectrics.

**Acknowledgements**

Discussions with V.Ya. Shur are gratefully acknowledged. S.V.K. research is partially supported by the U.S. Department of Energy, Basic Energy Sciences, Materials Sciences and Engineering Division.

# Appendix A

## A.1. Estimations of the depolarization field strength

Here we estimate the strength of the depolarization field (1b) caused by the interaction of uncharged domain wall with well-localized elastic fields in the ferroelectric material with free carriers. Let us consider a well-localized elastic field with characteristic z-size $R$, i.e. $\sigma_{ij}(\mathbf{r}) \equiv \sigma_{ij}^d(x,y)\theta(R-|z|)$. Note, that $\{x,y\}$ localization of elastic field does not affect the depolarization field (1b) acting on the component $P_z$. Since $\partial \sigma_{ij}(\mathbf{r})/\partial z \neq 0$, appearance of z-dependent coefficient $\alpha(T,x,y,z)$ in Eq.(1a) can lead to the depolarization field (1b), which value in the vicinity of defect can be estimated as $|E_d| \leq \dfrac{\delta P_S}{\varepsilon_0 \varepsilon_b}\exp(-R/R_D)$, where

$$\delta P_S = \left(\sqrt{-\frac{a(T)+2Q_{33ij}\sigma_{ij}^d}{\beta}} - \sqrt{-\frac{a(T)}{\beta}}\right) \approx P_S\left|\frac{Q_{33ij}\sigma_{ij}^d}{a(T)}\right|$$

($P_S$ is a spontaneous polarization). Using estimation for the ratio $\left|\dfrac{Q_{33ij}\sigma_{ij}^d}{a(T)}\right| \leq (10^{-2} - 10^{-1})$ at parameters $Q_{33ij} \sim 0.1$ m$^4$/C$^2$, $\beta_{ij}^d \sim$(30-3)eV, $N_d^0 \leq (10^{24} - 10^{25})$ m$^{-3}$ and $a \sim 5\,10^7$m/F at room temperature $T=300$ K, we get $|E_d| \leq 5\times(10^{-4}-10^{-5})P_S/(\varepsilon_0\varepsilon_b)$ for elastic field localized in z-region $R \geq 5R_D$. Thus $|E_d|$ becomes at least 3 orders smaller than the coercive field $\sim \dfrac{P_S}{\varepsilon_0 \varepsilon}$ ($\varepsilon \sim 100$). Using concentration of the free carriers $n \sim (10^{23} - 10^{25})$ m$^{-3}$ typical for ferroelectrics-semiconductors, we estimate $R_D = \sqrt{\varepsilon_0\varepsilon_b k_B T/e^2 n} \sim$(12 – 1.2)nm at $T=300$ K, $k_B=1.3807\times10^{-23}$ J/K and permittivity $\varepsilon_b \sim 10$. So, the depolarization field originated from the x-domain wall interaction with elastic fields localized in z-region (60 – 6) nm can be neglected. Note, that the sizes (20 – 2) nm are the typical sizes of elastic domain wall width at room temperature [34]. Despite the estimation $|E_d| \leq 5\times10^{-4}P_S/(\varepsilon_0\varepsilon_b)$ essentially overestimates the depolarization field in the region of x-domain wall, since $|P_z| \ll P_S$ in the region, it for sure allows us to neglect depolarization fields originating from elastic fields with z-sizes more than $5R_D$. Since depolarization field (1b) is sensitive only to z-dependencies of physical quantities in Eqs.(1a) and becomes negligible for elastic fields with z-sizes $R$ more than $5R_D$. Under these limitations the solution of 1D-problem, when all quantities effectively depend only on the distance x from the 180-degree domain wall,

located far from the sample boundaries, can explain qualitatively the main features of domain wall interaction with 3D-elastic defect fields in ferroelectric-semiconductors.

### A2. Free energy density

Free energy density has a form [29]

$$f_b = \frac{a(T)+2Q_{33ij}\sigma_{ij}}{2}P_z^2 + \frac{\beta}{4}P_z^4 + \frac{g}{2}\left(\frac{\partial P}{\partial x}\right)^2 - P_z E_0 - \frac{s_{ijkl}}{2}\sigma_{ij}\sigma_{kl} + \tilde{\beta}_{ij}^d \sigma_{ij}(N_d - \bar{N}_d) \quad (A.1)$$

Elastic strain obeys equation

$$u_{ij} = -\frac{\partial f_b}{\partial \sigma_{ij}} = s_{ijkl}\sigma_{kl} - \tilde{\beta}_{ij}^d(N_d - \bar{N}_d) - Q_{33ij}P_z^2 \quad (A.2)$$

Solution of the equation (A.2) along with mechanical equilibrium conditions $\partial \sigma_{ij}/\partial x_j = 0$ and compatibility relations leads to the following contribution of elastic defects into the stress tensor $\sigma_{11}^d(x) = 0$, $\sigma_{22}^d(x) = \sigma_{33}^d(x) = \beta_{ij}^d(N_d(x) - \bar{N}_d)$, where $\beta_{ij}^d = c_{ijkl}\tilde{\beta}_{kl}^d$. Substitution of the elastic solution in the free energy (A.1) and elementary transformations lead to

$$\tilde{f}_b = \left(1+\Delta\alpha\frac{N_d - \bar{N}_d}{N_d^0}\right)\frac{a(T)}{2}P_z^2 + \frac{\beta}{4}P_z^4 + \frac{g}{2}\left(\frac{\partial P}{\partial x}\right)^2 - P_z E_0 \quad (A.3)$$

Where $\Delta\alpha = \dfrac{2Q_{33ij}\beta_{ij}^d N_d^0}{a(T)}$. In order to obtain analytical results we put $\sigma_{ii}^d(x) \sim \cos(2\pi(x-x_0)/d)$.

### A3. Threshold electric field

Let us estimate the threshold electric field required for the domain wall motion using direct variational method for the free energy (A.3). *For the static case $E_0=0$, one-parametric trial function for polarization distribution was chosen as

$$P(x) = P_S\left(1+C\cos\left(\frac{2\pi(x-x_0)}{d}\right)\right)\tanh\left(\frac{x}{\sqrt{2}R_c}\right) \quad (A.4)$$

Where $C$ is a variational parameter. Very far from the wall (i.e. at $|x/R_c| \gg 1$) Eq.(A.3) gives $P(x) \to P_S(1+C\cos(2\pi(x-x_0)/d))$. Substitution of the trial function into the free energy

$$\langle f_b \rangle = \lim_{N\to\infty}\frac{1}{2Nd}\int_{-Nd}^{Nd}\left(f_b(x) - \frac{a(T)}{2}P_S^2 - \frac{\beta}{4}P_S^4\right)dx \quad (A.5)$$

after rather cumbersome but straightforward calculations gives $\langle f_b \rangle \approx \dfrac{a^2}{2\beta}\left(C^2\left(1-\dfrac{2g\pi^2}{ad^2}\right)-C\Delta\alpha\right)$

for $E_0=0$. Equation $\partial\langle f_b\rangle/\partial C = 0$ gives

$$C = \frac{\Delta\alpha \cdot d^2}{2(d^2 + 2\pi^2 R_C^2)} \quad (A.6)$$

Substitution of Eq.(A.6) into $\langle f_b \rangle$ gives $\langle f_b \rangle \approx -\frac{a^2}{2\beta}\left(1 - \frac{2g\pi^2}{ad^2}\right)C^2 \equiv -\frac{a^2}{2\beta}\left(1 + \frac{2\pi^2 R_C^2}{d^2}\right)C^2$.

For small $\Delta\alpha$ the polarization "ripples" are

$$\delta P = \frac{\Delta\alpha P_S \cdot d^2}{2(d^2 + 2\pi^2 R_C^2)} \cos\left(\frac{2\pi(x - x_0)}{d}\right) \tanh\left(\frac{x}{\sqrt{2}R_c}\right) \quad (A.6)$$

The ripples are shown in **Fig.A1b.** To move the domain wall over the ripples by a dragging field $E_0$ the latter should be overcame the threshold field. So that the threshold field can be estimated as averaged over the domain wall: $E_{th} \approx \frac{\langle \delta P \rangle|_{DW}}{\chi(T)}$, where the linear dielectric susceptibility $\chi(T) = \frac{1}{a + 3\beta P_S^2} \equiv \frac{1}{-2a(T)}$ was introduced.

$$E_{th} \approx \frac{3\sqrt{3}d^2 \Delta\alpha E_C}{2(d^2 + 2\pi^2 R_C^2)} \max_{x_0}\left[\frac{1}{2\sqrt{2}R_C}\int_{-\sqrt{2}R_c}^{\sqrt{2}R_c} \cos\left(\frac{2\pi(x - x_0)}{d}\right)\tanh\left(\frac{x}{\sqrt{2}R_C}\right)dx\right]$$

$$\approx \frac{3\sqrt{3}d^2 \Delta\alpha E_C}{2(d^2 + 2\pi^2 R_C^2)} \max_{x_0}\left[\frac{d}{8\pi^2 R_C^2}\left(-2\pi\sqrt{2}R_C \cos\left(\frac{2\pi\sqrt{2}R_c}{d}\right) + d\sin\left(\frac{2\pi\sqrt{2}R_c}{d}\right)\right)\sin\left(\frac{2\pi x_0}{d}\right)\right] \quad (A.7)$$

$$= \frac{3\sqrt{3}d^2 \Delta\alpha E_C}{2(d^2 + 2\pi^2 R_C^2)} \frac{\sqrt{2}d}{4\pi R_C}\left(-\cos\left(\frac{2\pi\sqrt{2}R_c}{d}\right) + \frac{d}{2\pi\sqrt{2}R_c}\sin\left(\frac{2\pi\sqrt{2}R_c}{d}\right)\right)$$

Pade approximation

$$E_{th} = \frac{3\sqrt{3}d^2 \Delta\alpha E_C}{2(d^2 + 2\pi^2 R_C^2)} \frac{\sqrt{2}d}{4\pi R_C}\left(-\cos\left(\frac{2\pi\sqrt{2}R_c}{d}\right) + \frac{d}{2\pi\sqrt{2}R_c}\sin\left(\frac{2\pi\sqrt{2}R_c}{d}\right)\right)$$

$$\approx \frac{3\sqrt{3}d^2 \Delta\alpha E_C}{2(d^2 + 2\pi^2 R_C^2)} \frac{R_C d}{2\sqrt{2}\pi R_C^2 - (5/4)R_C d + (3/2\sqrt{2}\pi)d^2} = \frac{3\sqrt{3}d^2 \Delta\alpha E_C}{2(d^2 + 2\pi^2 R_C^2)}\begin{cases}\frac{2\sqrt{2}\pi d}{8\pi^2 R_C}, & d \ll R_C \\ \frac{2\sqrt{2}\pi R_C}{3d}, & d \gg R_C\end{cases} \quad (A.8)$$

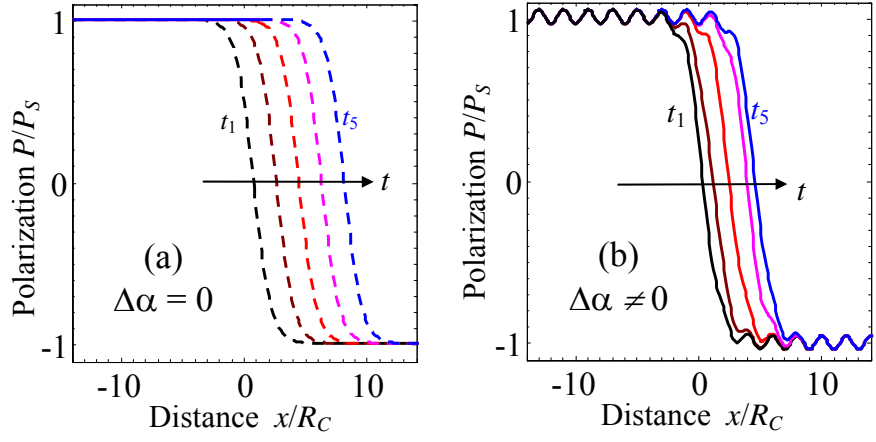

**Figure A1.** Temporal evolution of the polarization profile across the domain wall calculated without defects (a) and (b) with defects ( $d = 2R_c$ ) at different moments of time $t$: $t_1 < t_2 < t_3 < t_4 < t_5$. Dragging field $E_0$ is $0.05E_C$. Ripples shown in plot (b) appear due to defects.